\documentclass{ws-p8-50x6-00}

\begin{document}

\title{Non-Linear Evolution and Parton Distributions at Low $x$.}

\author{M. Lublinsky}

\address{Department of Physics, Technion -- Israeli Institute of Technology,
Haifa 32000, ISRAEL\\ 
E-mail: mal@techunix.technion.ac.il}


\maketitle

\abstracts{We suggest a new procedure  for extrapolating  the parton
distributions  from HERA  to much higher energies. The procedure
suggested consists of two steps. First,  we solve the non-linear evolution
equation. Second, we introduce a correcting function for which we  write a
DGLAP-type linear evolution equation. The nonlinear  equation is
solved numerically and estimates for the saturation scale, as well as
for  the gluon density at THERA and LHC energies  are  made.}

\section{Introduction}

In the present talk we present a new approach for extrapolation of the parton densities
to low $x$\cite{LGLM}. 
By low $x$ we mean values up to $10^{-7}$, which correspond to the  energy
range potentially covered by the THERA and LHC. 

A conventional approach to parton density evolution is based on the linear DGLAP 
equation which describes the gluon 
radiation leading to an increase in the number of 
partons. However, at  low $x$ the parton cascade become  
dense and the recombination processes 
start  to be important. We are convinced that such system cannot be described by a 
linear evolution any more. 

Our motivation is based on two main problems of the DGLAP 
evolution. First, it predicts a very steep growth of parton distributions at low $x$ violating 
the unitarity constraints. The second problem is  common for all perturbative series which
are asymptotic. In application to DIS processes it leads to  twist OPE break down
at low $x$,  when the high twists become of the same order as the leading one. We believe
that a non-linear evolution is a solution to both problems! The non-linear evolution accounts
for the saturation effects due to high parton density and it sums high twist contributions.

\section{New approach for extrapolating the parton distributions}

In the colour dipole picture  the evolution is applied to the imaginary part of
the dipole elastic scattering amplitude $N(r_\perp,x;b)$ for the dipole of the size $r_\perp$
elastically scattered at the impact parameter $b$. The amplitude $N$ is the major unknown
to be determined. Our approach to the problem is based on two steps in which we obtain
$N$ as a sum of two terms: $N=\tilde N+\Delta N$. As the first step, 
the function $\tilde N$ is found as a solution of the non-linear evolution 
equation~(\ref{EQ}).
This equation is valid in
the  leading $\ln(x)$ approximation of QCD. Moreover, it does not describe
correctly the evolution at  very short distances. In order to improve we perform the second
step of our program: the correcting function $\Delta N$ is introduced 
to incorporate the correct DGLAP kernel at short distances.  For $\Delta N$ 
we write down a DGLAP-type linear evolution equation. 

The non-linear evolution equation for the function $\tilde N$ has the form: 
\begin{eqnarray} \label{EQ}
-\frac{d 
\tilde{N}(r_{01},x)}{d\ln x}=
  \frac{C_F\,\alpha_s}{\pi^2}  
 \int d^2 r_{2}
\frac{r^2_{01}}{r^2_{02}
r^2_{12}} 
\left(2\tilde{N}(r_{02},x)-\tilde{N}(r_{02},x)
\tilde{N}(r_{12},x)\right) 
\end{eqnarray}
The above equation was derived in quite different approaches to high density QCD\cite{NLE}
and hence we believe it to be a very reliable tool. The equation~(\ref{EQ})  
 is derived for the large number of colors $N_c$ and for a  constant
$\alpha_s$.  We have dropped the $b$-dependence  implying the large $b$
limit. Having assumed $b$ to be large we will  allow to ourselves to extrapolate back to $b=0$.
The non-linear equation~(\ref{EQ}) is a subject to certain initial conditions, which we 
set at $x_0=10^{-2}$. The initial conditions are taken in the Glauber-Mueller form.

\section{Numerical solution}
\begin{figure}[t]
\begin{center}
\begin{tabular}{c  c}
\epsfxsize=11pc 
\epsfbox{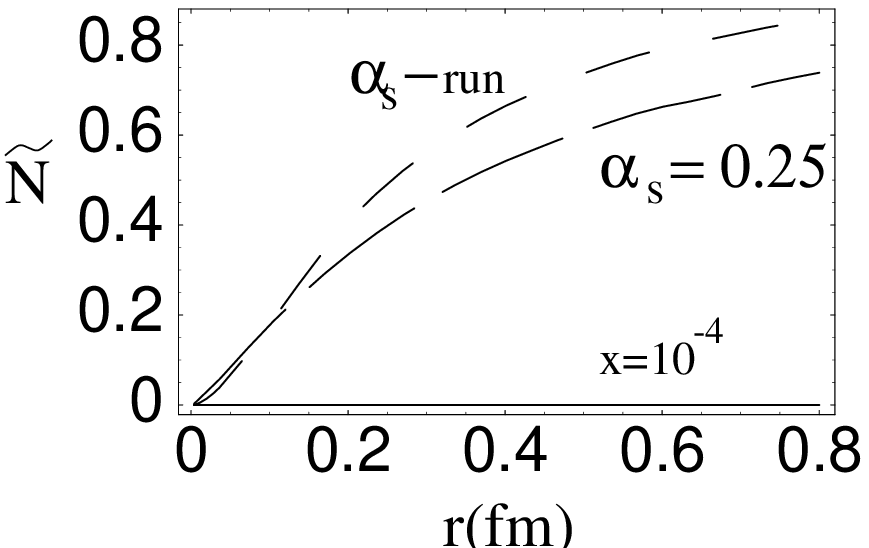} & 
\epsfxsize=10pc  \epsfbox{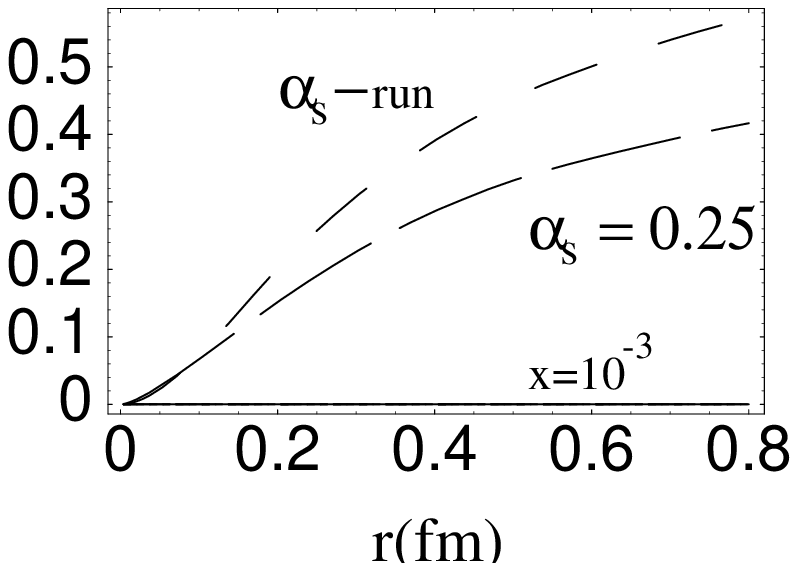} 
\end{tabular}
\caption{Solutions of the Eq.~(\ref{EQ}) as a function of distance.  \label{fig:sol}}
\end{center}
\end{figure}

The equation~(\ref{EQ}) is solved numerically by the method of iterations\cite{LGLM} for
 $b=0$. The 
solutions are obtained for the fixed $\alpha_s=0.25$ and  for the one loop running $\alpha_s$.
The figure~\ref{fig:sol} presents examples of the solutions as a function of  distance. The
solutions display a step like behavior consistent with all theoretical predictions. At short
distances the function $\tilde N$ tends to zero, while at large distances it approaches unity,
which is the unitarity bound.

\section{Results}

Basing on the above  solutions  several physical quantities can be determined.

\begin{figure}[t]
\begin{center}
\epsfxsize=10pc 
\epsfbox{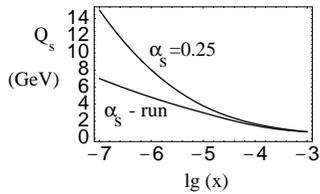} 
\begin{minipage}{7.0 cm}
\vspace*{-4cm}
\caption{Saturation scale as a function of $\lg x$.  \label{fig:satscal}}
\end{minipage}
\end{center}
\end{figure}

$\bullet$ The saturation scale $Q_s(x)$ is estimated as a typical scale where the function
$\tilde N$ undergoes the step like transition shown above. Though no exact mathematical
definition of the saturation scale is known we proposed several reasonable definitions
which in average produce our predictions depicted in the figure~\ref{fig:satscal}.

 $\bullet$ The solutions found display the scaling phenomena. Namely, the function
$\tilde N$ is not a function of two independent variables $x$ and $r_\perp$, but 
a function of a single variable $\tau=r_\perp Q_s(x)$:
$
\tilde N(x,r_\perp)\,\simeq\,\tilde N(r_\perp Q_s(x)).
$
The scaling holds with a few percent  accuracy in a  broad kinematic
domain below $x=10^{-2}$.

\begin{figure}[t]
\begin{center}
\begin{tabular}{c  c}
\epsfxsize=10pc 
\epsfbox{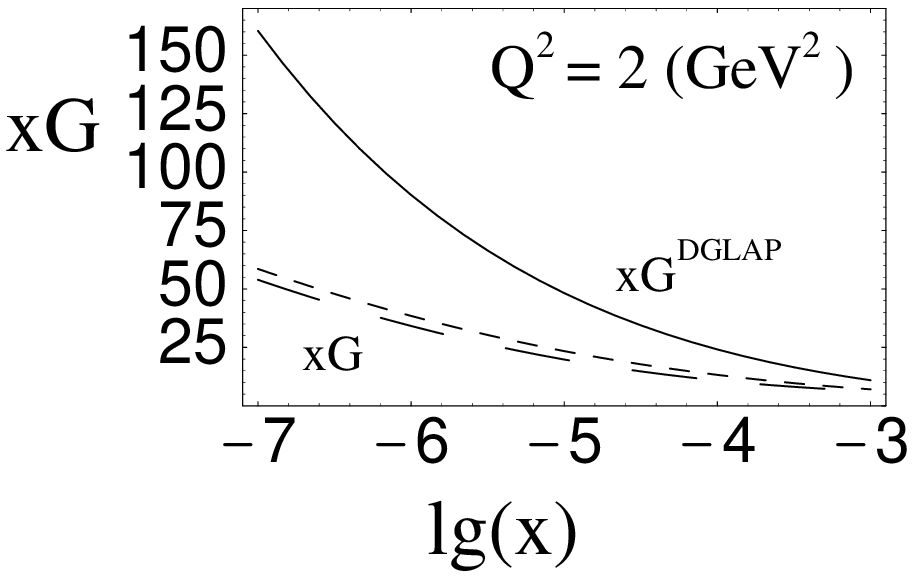} & 
\epsfxsize=10pc  \epsfbox{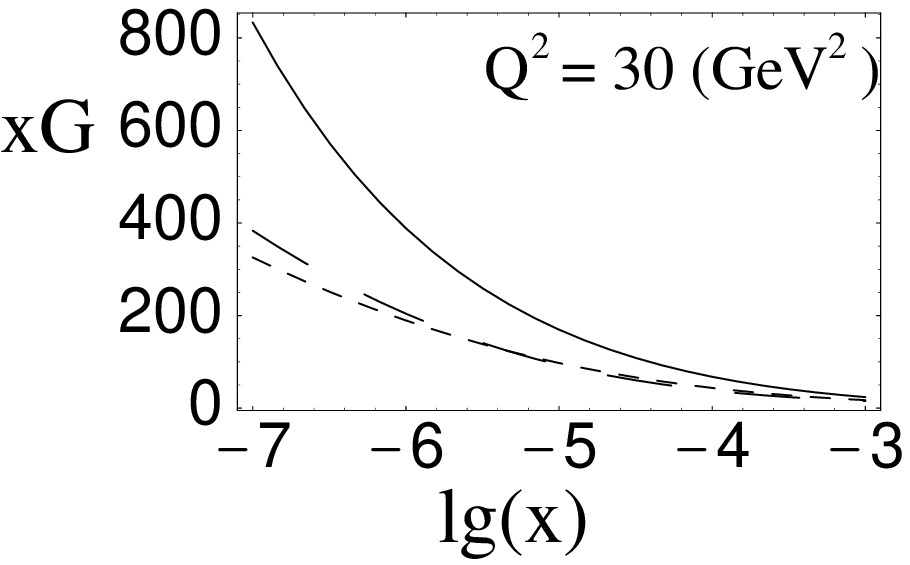} 
\end{tabular}
\caption{Gluon density as a function of $\lg x$.  \label{fig:density}}
\end{center}
\end{figure}

$\bullet$ The gluon density was defined according to the Mueller formula\cite{mu}, which 
relates the gluon density to the dipole elastic 
scattering amplitude. Examples of our predictions 
are shown in the Fig.~\ref{fig:density}. At small $x$ the gluon density is damped by a factor
2-3 comparing to the DGLAP predictions\cite{KKM}.

$\bullet$ The structure function $F_2$ is computed. The experimental data
below $x=10^{-2}$ is reproduced
with less than 20\% error  
 for $Q^2\le 50$ GeV$^2$. It is important  that the
data  is reproduced without use of the full DGLAP kernel.

\begin{figure}[t]
\begin{center}
\begin{tabular}{c  c}
\epsfxsize=11pc 
\epsfbox{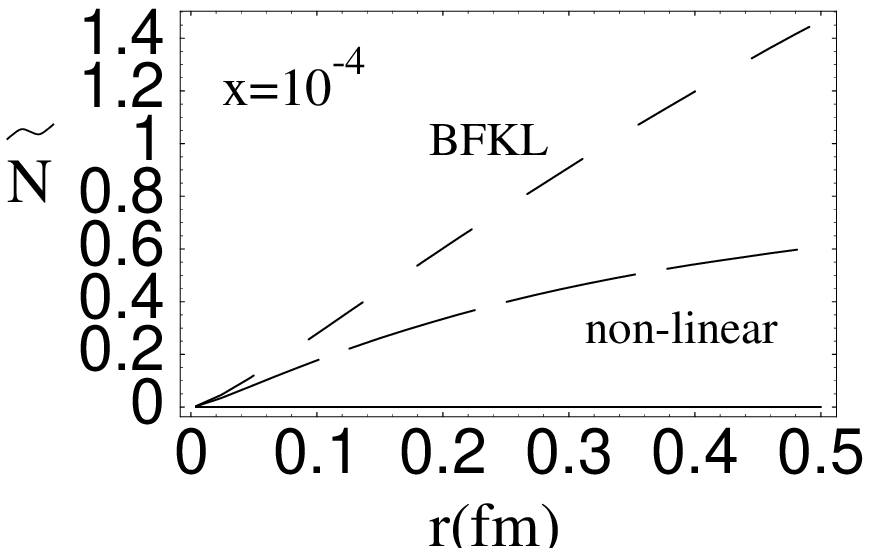} & 
\epsfxsize=10pc  \epsfbox{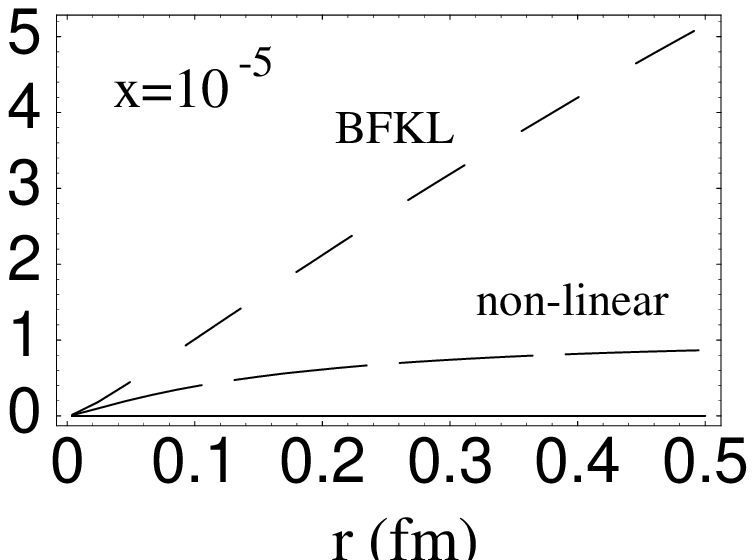} 
\end{tabular}
\caption{Comparison between the solutions of the LO BFKL equation and the Eq.~(\ref{EQ}).
 \label{bfkl}}
\end{center}
\end{figure}

$\bullet$
The LO BFKL equation, which is the linear part of the equation~(\ref{EQ}), is solved numerically
by the same method of iterations. The obtained results are compared in the Fig.~\ref{bfkl}
with the corresponding
solutions of the non-linear equation~(\ref{EQ}). The solution  of the BFKL equation  rapidly 
diverges from the solution of the non-linear equation. We conclude that the shadowing effects 
become important before the BFKL dynamics actually takes place.

\section{Summary and plans for the future}
A new method for extrapolating the parton densities to high energies was proposed. As the first
step in our program 
we solved (numerically) the non-linear evolution equation~(\ref{EQ}). From the solutions 
obtained we estimated  the saturation scale as a function of $x$. 
We found an approximate scaling behavior of the solutions. 
Both $xG$ and $F_2$ are found to be significantly damped at high energies.
We predicted  the LO BFKL dynamics  unlikely to be ever  seen.
We work currently on single diffractive dissociation from the non-linear evolution.
\vspace*{-0.5cm}
\section*{Acknowledgments}
I am especially grateful to  Eugene Levin for his cooperation  in the presented work.  
Part of the above is a  collaborating effort with  E. Gotsman and U. Maor and I 
wish to thank them for helpful discussions.

\end{document}